\documentclass[sigconf,natbib=true]{acmart}
\usepackage[ruled,linesnumbered]{algorithm2e}
\usepackage{multirow}
\usepackage{makecell}
\usepackage{enumitem}
\usepackage{diagbox}
\setitemize[1]{noitemsep,partopsep=0pt,parsep=0pt,topsep=0pt, leftmargin=10pt,
rightmargin=0pt}
\AtBeginDocument{%
  \providecommand\BibTeX{{%
    \normalfont B\kern-0.5em{\scshape i\kern-0.25em b}\kern-0.8em\TeX}}}

\setcopyright{acmcopyright}
\copyrightyear{2018}
\acmYear{2018}
\acmDOI{10.1145/1122445.1122456}

\acmConference[Woodstock '18]{Woodstock '18: ACM Symposium on Neural
  Gaze Detection}{June 03--05, 2018}{Woodstock, NY}
\acmBooktitle{Woodstock '18: ACM Symposium on Neural Gaze Detection,
  June 03--05, 2018, Woodstock, NY}
\acmPrice{15.00}
\acmISBN{978-1-4503-XXXX-X/18/06}

\begin{document}

\title{Distribution-Guided Auto-Encoder for User Multimodal Interest Cross Fusion}


\author{Moyu Zhang}
\affiliation{%
  \institution{Alibaba Group}
  \city{Beijing}
  \state{Beijing}
  \country{China}
}
\email{zhangmoyu@bupt.cn}

\author{Yongxiang Tang}
\affiliation{%
  \institution{Unaffiliated}
  \city{Beijing}
  \state{Beijing}
  \country{China}
}
\email{tangyongxiang94@gmail.com}

\author{Yujun Jin}
\affiliation{%
  \institution{Alibaba Group}
  \city{Beijing}
  \state{Beijing}
  \country{China}
}
\email{jinyujun.jyj@alibaba-inc.com}

\author{Jinxin Hu}
\authornote{Corresponding Author}
\affiliation{%
  \institution{Alibaba Group}
  \city{Beijing}
  \state{Beijing}
  \country{China}
}
\email{jinxin.hjx@alibaba-inc.com}

\author{Yu Zhang}
\affiliation{%
  \institution{Alibaba Group}
  \city{Beijing}
  \state{Beijing}
  \country{China}
}
\email{daoji@lazada.com}

\begin{abstract}
Traditional recommendation methods rely on correlating the embedding vectors of item IDs to capture implicit collaborative filtering signals between the user's historically clicked items and the target item, in order to model the user's interest in the target item. Consequently, traditional ID-based methods often encounter data sparsity problems stemming from the sparse nature of ID features. To alleviate the problem of item ID sparsity, recommendation models incorporate multimodal item information to enhance recommendation accuracy. However, existing multimodal recommendation methods typically employ early fusion approaches, which focus primarily on combining text and image features, while neglecting the contextual influence of user behavior sequences. This oversight prevents dynamic adaptation of multimodal interest representations based on behavioral patterns, consequently restricting the model's capacity to effectively capture user multimodal interests. Therefore, this paper proposes the \textbf{D}istribution-Guided \textbf{M}ultimodal-Interest \textbf{A}uto-\textbf{E}ncoder (DMAE), which achieves the cross fusion of user multimodal interest at the behavioral level. Specifically, DMAE comprises three key components: 1) Multimodal Interest Encoding Unit (MIEU), which encodes the similarity scores between the target item and historically clicked items as the corresponding representation vectors of user interest across different modalities. 2) Multimodal Interest Fusion Unit (MIFU), which dynamically adapts user interest representations derived from user behavior sequences across modalities via intra- and inter-modal cross-fusion, enabling fine-grained multimodal interest fusion with awareness of the behavioral context. 3) Interest-Distribution Decoding Unit (IDDU), which employs a decoder to reconstruct the encoded user interest representations into true similarity distributions for each modality. The similarity distributions serve as a guide for model learning, aiming to retain as much multimodal information as possible. Ultimately, extensive experiments demonstrate the superiority of DMAE.
\end{abstract}


\maketitle

\section{Introduction}
In the current age of information overload, recommendation systems (RS) play a crucial role in assisting users in identifying items of interest from an overwhelming array of options \cite{rs1, rs2, rec1, rec2}.  By predicting the probability of users clicking on a specific item, we can prioritize the target item set for each user and personalize the display of the most appealing items to users. Up to now, recommendation models have garnered significant interest from both research and industry in recent years and have found widespread use in various online service scenarios, including recommendation and pay-per-click (PPC) advertising systems  \cite{din, dien}.

\begin{figure*}[t]
  \centering
  \includegraphics[width=\linewidth]{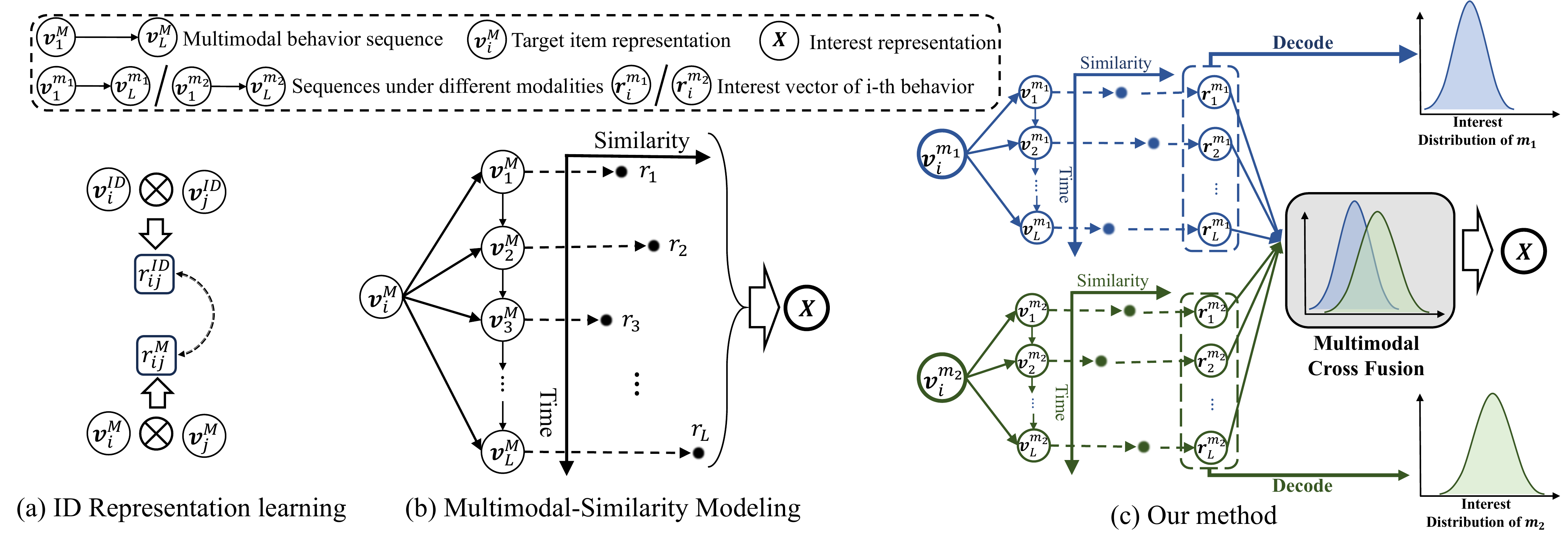}
  \caption{Examples of previous methods and ours. In (a), the representation learning of ID is optimized by constraining $r_{ij}^{ID}$ and $r_{ij}^{M}$ to be close. While in (b), the similarity is calculated with the early fused multimodal representation, and the intra-modal crossover cannot be effectively achieved. In (c), we achieve intra-modal and inter-modal crossover through multimodal cross fusion at the behavior granularity. We also use the decoding process to avoid the loss of multimodal information.}
  \label{example_method}
\end{figure*}  

However, in non-search recommendation scenarios, users often do not explicitly express their intentions. Thus, the recommendation model needs to analyze the user's interest in the target item based on their historical click sequence in order to accurately predict the probability of the user clicking on the target item and to make appropriate recommendations. With the advancement of deep learning technology in recent years, current recommendation models typically compress user sequences into one or more hidden representation vectors that represent user interests, simplifying the prediction of user clicks on the target item \cite{sim, gru4rec}. Among these models, the most popular method is the attention-based sequence modeling method \cite{din, dien, bst}, which captures the implicit collaborative filtering signal between the target item and the historically clicked items through the embedding vector corresponding to the item ID. This method adaptively learns the user interest representation according to the user's historical behaviors.

Despite the considerable success of previous sequence modeling methods, most current industrial systems still rely exclusively on sparse item IDs to model user interests. As a result, recommendation systems often encounter the challenge of inadequate item representation learning in practice due to data sparsity \cite{rd}, impacting the accuracy of recommendations. Furthermore, the item ID embedding vector learned by the recommendation model typically captures co-occurrence information between items, such as users clicking on item A and then on item B, but cannot effectively capture the explicit semantic information in the item text and image. In recent years, with the successful adoption of Large Language Models (LLMs) across various fields, the recommendation field has increasingly sought to incorporate multimodal information of items into recommendation models to leverage the abundant world knowledge of LLMs \cite{llm1, llm2, llm3}. Nonetheless, because of the limitations in online computing resources and constraints in response time, online recommendation models typically cannot directly utilize the high-dimensional multimodal representations generated by LLMs. This is due to the fact that the multimodal representations are often in a much higher dimension than the embedding vector dimensions of the ID and are challenging to jointly train based on sparse recommendation samples. Hence, in current recommendation field, the predominant approach for leveraging multimodal information involves freezing the multimodal representations and utilizing the similarity relationship between the items as auxiliary features to aid the model in capturing the user's interest in the target item.

Currently, mainstream multimodal methods can be categorized into two types: 1) relationship-guided ID representation learning, where multimodal similarity between items is used as a training label in a pair-wise training task, integrating multimodal information into the item ID representation \cite{lrd}. When serving online, only the ID representation is used to model user interests, as shown in Figure \ref{example_method}(a). However, this method may result in some loss of multimodal information during the training process. 2) multimodal-similarity modeling, which keeps the item ID representation and multimodal representation independent from each other \cite{scl}, as shown in Figure \ref{example_method}(b). It introduces the multimodal similarity sequence between the target item and historically clicked items as additional information to model user interest. This method retains multimodal information to the greatest extent without affecting the learning of co-occurrence. However, both methods are considered early multimodal fusion, as they use single representation to represent multimodal information of items, but fail to adaptively adjust the user interests reflected in different modalities according to the user behavior sequence. Considering the importance of the user behavior sequence in modeling user interests in recommendation scenarios, fine-grained behavioral-level multimodal interest fusion enables deeper mining of item multimodal information, consequently enhancing recommendation accuracy.

Therefore, to effectively capture user multimodal interest, we introduce the \textbf{D}istribution-Guided \textbf{M}ultimodal-Interest \textbf{A}uto-\textbf{E}ncoder (DMAE) to dynamically adapts user interest representations at the behavioral level, as shown in Figure \ref{example_method}(c). Specifically, DMAE consists of three main components: (i) Multimodal Interest Encoding Unit (MIEU), which uses item representation vectors from different modalities to calculate similarity scores between the target item and historically clicked items, resulting in various item similarity score sequences across various modalities. Through an encoder, it combines similarity scores with the time of click behavior, and maps them into a multidimensional vector representing user interest reflected by the corresponding click behavior. (ii) Multimodal Interest Fusion Unit (MIFU), which first adapts user interest representations within each modality to facilitate intra-modal information transfer. Subsequently, it incorporates single-modality user interests as contextual guidance for cross-modal interest fusion, achieving comprehensive multimodal integration that combines both intra-modal and inter-modal perception. (iii) Interest-Distribution Decoding Unit (IDDU), which reconstructs the encoded user interest representations into true similarity distributions for each modality \cite{mae}. The similarity distributions serve as a guide for model learning, aiming to retain as much multimodal information as possible. 

The contributions of our paper can be summarized as follows:
\begin{itemize}
\item To our knowledge, DMAE is the first model that dynamically fuses users’ multimodal interests based on their historical sequences to make full use of multimodal information of items.
\item DMAE uses an encoder to map user interests reflected by user sequences across different modalities, and achieves multimodal interest cross fusion of behavior granularity by conducting intra-modal and inter-modal interactions.
\item DMAE introduces a decoder that utilizes user interest distribution as a label to guide the learning of model encoder parameters in order to minimize the loss of multimodal information during the training process of the recommendation task.
\item Evaluations using offline datasets and online A/B testing have demonstrated the superiority of the proposed DMAE method.
\end{itemize}  
 
\section{Related Work}
\subsection{Traditional Interest Modeling}
In recent years, deep learning has propelled the recommendation model from a linear approach to a deep network-based approach, which includes Wide \& Deep \cite{widedeep}, DeepFM \cite{deepfm}, PNN \cite{pnn}, and other methods \cite{xdeepfm, can, autoint}. However, these methods primarily focus on feature interaction and do not effectively mine user interest reflected in user historical behavior. Previous user sequence modeling methods can be categorized into four types: traditional pooling methods \cite{rec1}, RNN-based methods \cite{gru4rec, dien, dupn, dhan}, capsule-based methods \cite{comirec, mind, mgnm}, and attention-based methods \cite{din, dstn, atrank, bst, dsin, tissa, sdm, kfatt, dfn, sim, dmt, mimn, pinnerformer}. Pooling-based methods use average, sum, or max pooling to merge user behavior sequences \cite{rec1}, treating all behaviors in a user's history sequence as equally important, thus reflecting the user's primary interests. To capture the time-series features of user sequences, researchers developed RNN-based models, such as GRU4Rec \cite{gru4rec}, DUPN\cite{dupn}, DHAN \cite{dhan}, etc. Later, to enrich recommendation diversity, researchers introduced capsule network-based models, such as MIND\cite{mind} and ComiRec \cite{comirec}, to model user sequences using Capsules Network \cite{capsule} and implicitly cluster user sequences to obtain multiple interest vectors, allowing them to learn from multiple categories of items being recalled. Regardless of whether Capsules, RNN, or pooling-based methods are used, they model all of a user's interests rather than precise interests in the target item. As a result, when sorting different target items for the CTR task, the user's interest vector remains constant, which can overshadow their interest in target items. To accurately model users' interest in the target item, researchers have proposed attention-based methods, such as the target-attention mechanism, which have become popular recently. For instance, DIN \cite{din} and DIEN \cite{dien} models relevant behavior for recommending items. 
\subsection{Multimodal Interest Modeling}
Despite the excellent performance of traditional methods, they typically rely on sparse item IDs to extract user interests.  The data sparsity issue in recommendation systems often leads to inadequate item ID representation learning in recommendation models, impacting recommendation accuracy \cite{rd}. Furthermore, the item ID embedding vectors learned by the recommendation model typically implicitly capture co-occurrence information between items. For instance, when a user clicks on item A and then on item B, the model typically allows the ID embeddings of items A and B to reflect this correlation, but cannot effectively capture the explicit semantic information from the text and image of items. In contrast, multimodal data can explicitly offer richer semantic information. In recent years, the successful application of Large Language Models (LLMs) in various fields has also laid the foundation for the use of multimodal data in the recommendation field. Current mainstream approaches to using multimodal information are twofold: 1) relationship-guided ID representation learning, where multimodal similarity between items is used as a training label in a pair-wise training task, integrating multimodal information into the item ID representation \cite{lrd}. However, this method may result in loss of multimodal information during the training process. 2) multimodal-similarity modeling, which keeps the item ID representation and multimodal representation independent from each other \cite{scl}. It introduces the multimodal similarity sequence between the target item and historically clicked items as additional information to model user interest. However, both methods utilize early fused multimodal representations and  do not adaptively adjust the user interests reflected by different modalities based on users' behaviors, thus failing to achieve intra-modal and inter-modal interactions. 

\section{Preliminaries}
This section begins by formulating the recommendation task, followed by presenting the fundamental inputs of our model.

\subsection{Problem Formulation}
The recommendation task is a binary classification problem that deals with sparse multi-domain feature data, typically composed of three types of features: \emph{User Profile} $x_u$, \emph{User Behavior Sequence} $\mathcal{S}$, and \emph{Target Item} $x_i$, which collectively form the input feature set $\mathcal{X}$. Each prediction instance is denoted as $(\mathcal{X}, y)$, where $y \in \left\{0, 1 \right\}$ (click or no-click) is the label of user behavior. The user behavior sequence usually comprises multiple items that users have clicked in the past, which can be represented by $\mathcal{S} = \left\{ x_1, ..., x_j, ..., x_T \right\}$. Here, $T$ denotes the number of historical user behaviors. Then, the prediction task aims to approximate the probability $P(y|\mathcal{X})$.

\subsection{Fundamental Inputs}
Currently, most of recommendation methods follow the paradigm of \emph{Input $\rightarrow$ Embedding $\rightarrow$ Prediction}. The traditional embedding layer involves the model randomly initializing an embedding vector table for each feature field and optimizing the parameters of these embedding vectors during the training task. Building on this, we also introduce additional multimodal representations for items: \\
 \textbf{$\bullet$ ID Embedding}. As the input features are typically high-dimensional one-hot vectors, we incorporate an embedding layer to transform $x_u$ and $x_i$ into low-dimensional vectors $\boldsymbol{v}_u$ and $\boldsymbol{v}_i$, respectively. Each historically clicked item $x_j$ in $\mathcal{S}$ can be embedded as $\boldsymbol{v}_j$. \\
  \textbf{$\bullet$ Multimodal Embedding}. Typically, in practical settings, fully modeling the explicit semantic association between items based solely on their exposure records in the recommendation system is not feasible.  \textbf{\emph{Images}} or  \textbf{\emph{text descriptions}} of items can often provide an intuitive sense of the relationship between two items. Thus, we utilize LLMs for extracting text and image representations of items as multimodal information, integrating them into the models to improve recommendation accuracy. Each historically clicked  item $x_j$ in $\mathcal{S}$ can be embedded as $\boldsymbol{v}_j^{m_1}$ and $\boldsymbol{v}_j^{m_2}$, representing the text representation and image representation of the item, respectively. Similarly, the multimodal representation of the target item $x_i$ can be $\boldsymbol{v}_i^{m_1}$ and $\boldsymbol{v}_i^{m_2}$. The multimodal representation vector parameters are frozen during the recommendation model training process.

\begin{figure*}[t]
  \centering
  \includegraphics[width=\linewidth]{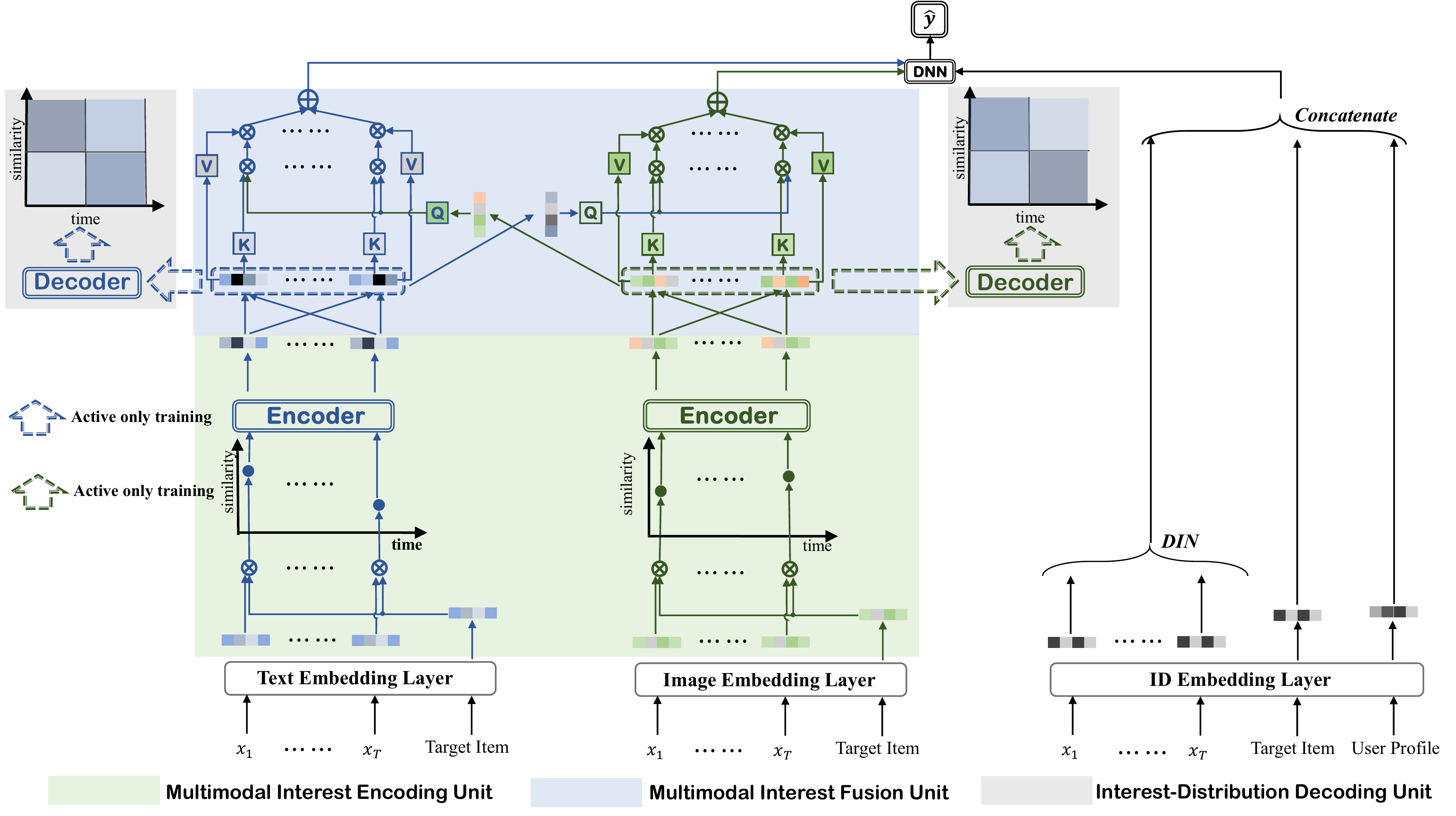}
  \caption{The structure of Distribution-Guided Multimodal-Interest Auto-Encoder (DMAE), where the Interest-Distribution Decoding Unit only participates in the model training and is not required during the test or online inference process.}
  \label{model}
  \vspace{-0.3cm}
\end{figure*} 

\section{Method}
User's historical click sequences reflect their interest, making it valuable to explore their interest in the target item based on the sequences for recommendations. In recent years, LLMs have led to an increase in the utilization of multimodal data in recommendation models, where the user's interest is determined by assessing the multimodal similarity between historically clicked items and the target item. However, previous methods often use early-fusion multimodal representations, ignoring information interactions within and between modalities of the users' sequence. To fully utilize the multimodal information, we propose a new approach called Distribution-Guided Multimodal-Interest Auto-Encoder (DMAE), which achieves multimodal interest fusion based on the user's sequence by enabling multimodal interest interactions at the behavioral granularity. Furthermore, to minimizes information loss during interest encoding, we propose constructing user interest distributions based on similarity scores between the target item and historically clicked items under different modalities as supervision signals to guide parameter learning. Figure \ref{model} illustrates the overall structure of DMAE, which consists of three key components: \\
\textbf{$\bullet$ Multimodal Interest Encoding Unit (MIEU)}, which combines item similarity scores across different modalities with the timing of click behavior, and then maps them into multidimensional vectors that can reflect the user's interest in the target item. \\
\textbf{$\bullet$ Multimodal Interest Fusion Unit (MIFU)}, which utilizes cross-modal and intra-modal information to redefine the interest representations of each behavior to achieve multimodal cross fusion. \\
\textbf{$\bullet$ Interest-Distribution Decoding Unit (IDDU)}, which decodes the interest representations into corresponding user interest distribution under different modalities to guiding the parameter learning.

\subsection{Multimodal Interest Encoding Unit (MIEU)}  
In this paper, we aim to model the user's interest in a target item within a representation space defined by the multimodal similarity of items. To accomplish this, we initially encode the similarity scores of historically clicked items and the target item into a multidimensional vector. Additionally, taking into account that the timing of a click may have varying effects on user interest, we integrate the position feature of the click behavior with the representation of the similarity score to derive the user interest representation vector corresponding to the behavior in each modality .

\subsubsection{Similarity Score Representation}
We calculate the similarity scores between the historically clicked items in the user sequence and the target item in each modality, resulting in two similarity score sequences $r^{m_1}= [r_1^{m_1}, ..., r_j^{m_1}, ..., r_T^{m_1} ]$ and $r^{m_2}= [r_1^{m_2}, ..., r_j^{m_2}, ..., r_T^{m_2} ]$, where $r_j^{m_1}=(\frac{\boldsymbol{v}_i^{m_1} \boldsymbol{v}_j^{m_1}}{|\boldsymbol{v}_i^{m_1}| |\boldsymbol{v}_j^{m_1}|}+1)/2 \in [0,1]$. Since similarity scores between items are numerical features, they cannot be directly embedded like sparse ID features. Previous CTR prediction models discretize continuous-valued features by setting intervals and then perform embedding operations \cite{deepfm}. However, hard discretization removes the differences between values within the same interval and the similarity between values on either side of the boundary. To accurately capture multimodal interest information, we employ discretization and sine-cosine encoding to jointly represent similarity scores, preserving differences at the vector level. This ensures a comprehensive representation of the multimodal similarity information. The specific process is as follows:

\textbf{\emph{Discretization}}. Assuming an input similarity score is given as $r$, Most current methods for numerical features involve discretizing values into different ranges by dividing boundaries \cite{deepfm}. This method significantly enhances the representation ability of numerical features by using multi-dimensional vectors as representations, compared with direct input. Since this method can achieve end-to-end training, it is more conducive to model learning. However, this approach loses the distinction between values in the same range. To address this, we multiply the discretized embedding vectors for each value by the scaled values of the model. This way, even if two values are discretized into the same embedding vector, we can add a certain degree of difference between the two vectors through the multiplication operation. The calculation is as follows:
\begin{gather}
\hat{r} = wr +b \\
\boldsymbol{r}^d = \hat{r} (Bucket(|r(N_B-1)|))
\end{gather} 
where $\hat{r}$ is the scaled value. $w \in \mathbb{R}^{1}$ and $b \in \mathbb{R}^{1}$ are learnable parameters, which are used to scale the similarity value to an appropriate range in an adaptive manner, thereby accelerating model convergence.  $|\cdot|$ stands for the rounding down operation. $Bucket(\cdot)$ denotes that the integer $|r(N_B-1)|$ is embed into corresponding bucket vector. $N_B$ is the number of bucket.  At last, we get $\boldsymbol{r}^d \in \mathbb{R}^{d}$ as the representation vector of $r$, where $d$ is the vector dimension.

\textbf{\emph{Sine-Cosine}}. The aforementioned discretization method only retains the differences between values at the vector-level, where each element of the vector is multiplied by a scaled value. To further preserve the differences between different similarity scores, we employ a method similar to the Transformer's sine-cosine position encoding \cite{transformer} to generate a $d$-dimensional representation vector for each similarity score, thereby distinguishing the differences between values at the element-level. The process is as follows:
\begin{gather}
\boldsymbol{r}^{sc} = (sin(\frac{r}{10^{\frac{2\times0}{d}}}), cos(\frac{r}{10^{\frac{2\times0}{d}}}), ..., sin(\frac{r}{10^{\frac{d}{d}}}), cos(\frac{r}{10^{\frac{d}{d}}}))
\end{gather} 

\textbf{\emph{Interest Mapping}}. We combine the above two representation vectors of each similarity score as the final representation vector for that score. At the same time, since user interests change over time, we need to combine similarity scores with position information to capture the dynamic changes in user interests. To achieve this, we introduce position-corresponding embedding vectors to consider the impact of the similarity between clicked items and the target item at different moments on user interests as follows:
\begin{gather}
\boldsymbol{r} = ReLU(\boldsymbol{W}_2 (ReLU(\boldsymbol{W}_1 (\boldsymbol{r}_{j}^{d} \oplus \boldsymbol{r}_{j}^{sc}) + \boldsymbol{b}_1) \oplus \boldsymbol{PE}_{T-j}) + \boldsymbol{b}_2)
\end{gather} 
where $\boldsymbol{W}_1 \in \mathbb{R}^{d \times 2d}$, $\boldsymbol{W}_2 \in \mathbb{R}^{d \times 2d}$,  $\boldsymbol{b}_1 \in \mathbb{R}^{d}$, and $\boldsymbol{b}_2 \in \mathbb{R}^{d}$ are learnable parameters. Using the above embedding method, the two similarity score sequence can be encoded as $\boldsymbol{r}^{m_1}=[\boldsymbol{r}_{1}^{m_1}, ...,\boldsymbol{r}_{j}^{m_1}, ..., \boldsymbol{r}_{T}^{m_1}]$, and $\boldsymbol{r}^{m_2}=[\boldsymbol{r}_{1}^{m_2},...,  \boldsymbol{r}_{j}^{m_2}, ..., \boldsymbol{r}_{T}^{m_2}]$, where the model parameters of embedding similarity score under different modalities are different. 

\begin{figure}[t]
 \centering
\includegraphics[width=\linewidth]{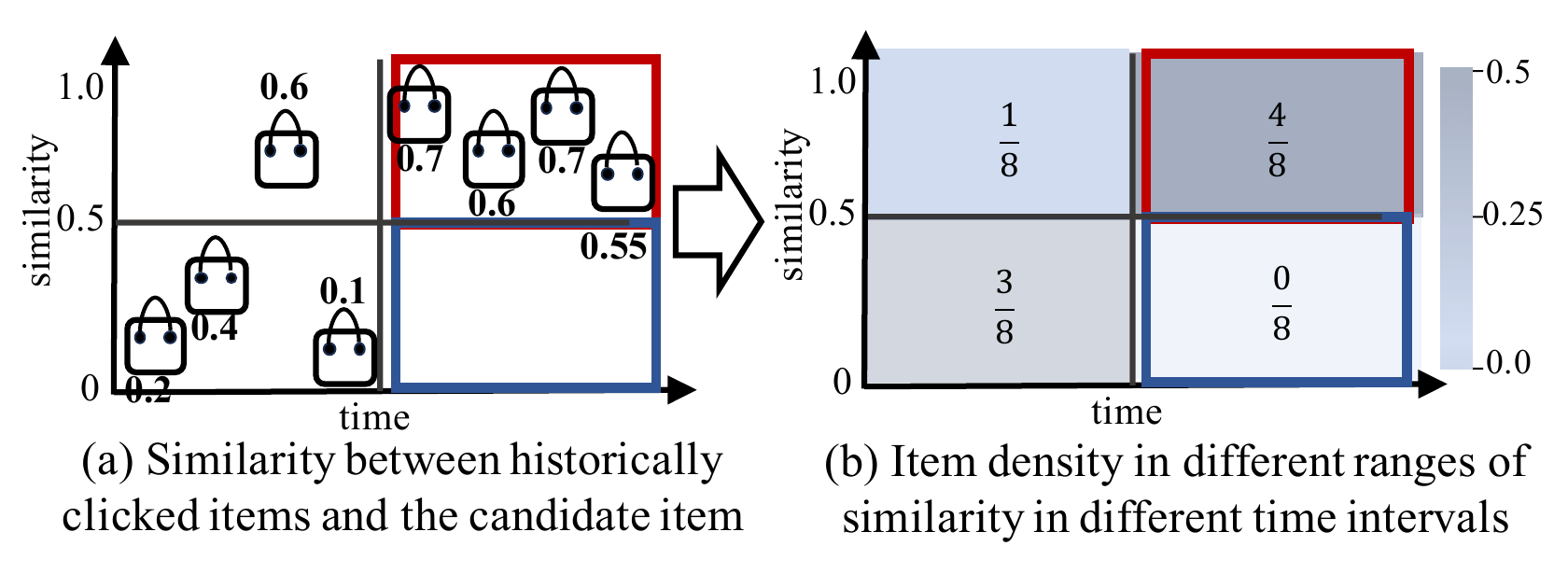}
\caption{Illustration of the interest distribution. In (b), the color of each area represents the proportion of items clicked in that area compared to the total click sequence. }
\label{interest_example}
\end{figure}

\subsection{Multimodal Interest Fusion Unit (MIFU)} 
As mentioned before, previous multimodal recommendation approaches typically employ early-fused representations to model item relationships, neglecting cross-modal interactions and particularly the influence of behavioral sequences on multimodal fusion. To enable behavior-aware multimodal fusion, we first process intra-modal interest representations through self-attention for intra-modal information exchange. We then integrate per-modality interests as contextual guidance for cross-modal fusion, enabling comprehensive multimodal integration. To avoid the significant increase in computational complexity, we use sliding window attention instead of the traditional self-attention mechanism:
\begin{gather}
[\hat{\boldsymbol{r}}_{1}^{m_1}, ..., \hat{\boldsymbol{r}}_{j}^{m_1}, ..., \hat{\boldsymbol{r}}_{T}^{m_1} ] = \mathcal{G}^{m_1}([\boldsymbol{r}_{1}^{m_1}, ..., \boldsymbol{r}_{j}^{m_1}, ..., \boldsymbol{r}_{T}^{m_1} ]) \\
[\hat{\boldsymbol{r}}_{1}^{m_2}, ..., \hat{\boldsymbol{r}}_{j}^{m_2}, ..., \hat{\boldsymbol{r}}_{T}^{m_2} ] = \mathcal{G}^{m_2}([\boldsymbol{r}_{1}^{m_2}, ...,  \boldsymbol{r}_{j}^{m_2}, ..., \boldsymbol{r}_{T}^{m_2} ])
\end{gather} 
where $\mathcal{G}^{m_1}(\cdot)$ and $\mathcal{G}^{m_2}(\cdot)$ denote the sliding window functions for the modalities. After that, to further extract cross-modal information, we aggregate the adjusted interest representation sequences from the two modalities and utilize them as contextual guidance for fusing the interest representation of each behavior across different modalities, ultimately achieving multimodal cross fusion. The detailed calculation process is as follows:
\begin{gather}
\overline{\boldsymbol{r}}^{m_1} = \frac{1}{T}\begin{matrix} \sum_{j=1}^T \hat{\boldsymbol{r}}_{j}^{m_1}   \end{matrix},  \quad \overline{\boldsymbol{r}}^{m_2} = \frac{1}{T}\begin{matrix} \sum_{j=1}^T \hat{\boldsymbol{r}}_{j}^{m_2}   \end{matrix}
\\
\boldsymbol{r}_{*}^{m_1} = softmax(\frac{(\overline{\boldsymbol{r}}^{m_2} \boldsymbol{W}_Q^{m_1})(\hat{\boldsymbol{r}}^{m_1} \boldsymbol{W}_K^{m_1})^T}{\sqrt{d}})\hat{\boldsymbol{r}}^{m_1} \boldsymbol{W}_V^{m_1}\\
\boldsymbol{r}_{*}^{m_2} = softmax(\frac{(\overline{\boldsymbol{r}}^{m_1} \boldsymbol{W}_Q^{m_2})(\hat{\boldsymbol{r}}^{m_2} \boldsymbol{W}_K^{m_2})^T}{\sqrt{d}})\hat{\boldsymbol{r}}^{m_2} \boldsymbol{W}_V^{m_2}
\end{gather} 
where $\overline{\boldsymbol{r}}^{m_1}$ and $\overline{\boldsymbol{r}}^{m_2}$ are the average pooling vectors of two multimodal interest sequence. $\left\{\boldsymbol{W}_Q^{m_1}, \boldsymbol{W}_K^{m_1}, \boldsymbol{W}_V^{m_1} \right\}$ and $\left\{\boldsymbol{W}_Q^{m_2}, \boldsymbol{W}_K^{m_2}, \boldsymbol{W}_V^{m_2} \right\}$ are the projection matrices for the modality $m_1$ and $m_2$, respectively.  $\hat{\boldsymbol{r}}^{m_1}=[\hat{\boldsymbol{r}}_{1}^{m_1}, \hat{\boldsymbol{r}}_{2}^{m_1}, ..., \hat{\boldsymbol{r}}_{T}^{m_1}]$ and $\hat{\boldsymbol{r}}^{m_2}=[\hat{\boldsymbol{r}}_{1}^{m_2}, \hat{\boldsymbol{r}}_{2}^{m_2}, ..., \hat{\boldsymbol{r}}_{T}^{m_2}]$. By applying the cross-attention method described above, we have redefined the interest representation for each modality in the user sequence. The pooled vector contains information of all behaviors within the modality sequence, allowing us to achieve intra-modal crossover using this method. Ultimately, we obtain two redefined multi-modal interest representation vector,  i.e., $\boldsymbol{r}_{*}^{m_1}, \boldsymbol{r}_{*}^{m_2} \in \mathbb{R}^{d}$, which already encompass both intra-modal and inter-modal crossover information.

\subsection{Interest-Distribution Decoding Unit (IDDU)} 
Even though user interests are modeled in the item similarity representation space, guaranteeing that the encoded interest representation vectors accurately reflect the true similarity distribution based solely on predicting users whether click the target item remains a challenge. Therefore, we introduce the concept of \textbf{\emph{Interest Distribution}} in this paper and propose to extract the user's interest distribution as labels of the decoding uint. Generally speaking, the historical clicks of a user on items highly similar to the target item typically signify a strong interest in the target item by the user. Additionally, the user's click behavior occurring at different times will impact the intensity of reflected interest, as interests tend to evolve over time. Therefore, as illustrated in Figure \ref{interest_example}(a), we partition the user's historical click sequence into several regions based on the similarity score and the time of occurrence. The proportion of click behaviors in these regions represents the user's interest distribution, as demonstrated in Figure \ref{interest_example}(b). A higher proportion of the upper right region (red outline) indicates a greater interest in the target item, as it suggests frequent recent clicks on items similar to the target item. Conversely, a high proportion of the lower right region (blue box area) signifies a lesser interest in the target item. 

Building upon this concept, we design a decoder structure to decode the interest representations in MIFU as the interest distribution and make the decoded distribution close to the real similarity distribution of the user sequence, so as to constrain the learning of encoding unit parameters to reduce the loss of information. 

\subsubsection{Construct Interest-Distribution}
We first divide the user sequence equally along the time dimension from $0$ to $T$ to generate $l$ subsequences. Next, we divide the range of similarity values from $0$ to $1$ equally into $n$ similarity range intervals, denoted as $c_1=[0, \frac{1}{n}]$, $c_2=(\frac{1}{n}, \frac{2}{n}]$, ..., and $c_n=[\frac{n-1}{n}, 1]$. Then, we calculate the ratio of the number of items with similarity scores in different intervals to the sequence length in each time range as follows:
\begin{gather}
p_{ij} = \frac{\# \left\{s_k | ( s_k \in c_j ) \& ( s_k \in \boldsymbol{s}_i ) \right\}}{ T}
\end{gather} 
where $p_{ij}$ denotes the proportion of the number of similarity scores between items in subsequence $s_i$ and the target item within the range of $c_j$ to the entire sequence. $\#\left\{ \cdot \right\}$ represents the number of instances that satisfy a certain condition. If most of the items clicked by the user in a certain subsequence have a high similarity with the target item, it indicates that the user has a high interest in the target item during this time period. Using the above method, we can construct different user interest distributions based on the user sequence under different modalities, denoted as $\boldsymbol{P}^{m_1}=[p_{ij}^{m_1}]_{1 \leq i \leq l}^{1 \leq j \leq n} \in \mathbb{R}^{l\times n}$ and $\boldsymbol{P}^{m_2}=[p_{ij}^{m_2}]_{1 \leq i \leq l}^{1 \leq j \leq n} \in \mathbb{R}^{l\times n}$, respectively.

\subsubsection{Decoder Module}  
To ensure that the interest representations output by the encoding unit can restore the true user interest distribution, we introduce a decoder structure that uses the constructed interest distributions as labels for decoding interest representations. Using the decoding loss for parameter optimization enables the encoder to minimize the loss of multimodal information. It is worth noting that our decoder module only participates in the calculation during offline training. During online inference, only the parameters of the encoder can be used for prediction. 

Specifically, in the decoding process, we add a random mask to the similarity score sequence to improve the model's generalization ability \cite{mae}. This is equivalent to sampling the user sequence, and then using a Multi-Layer Perceptron (MLP) to decode and construct the user interest distribution by predicting values in the range of 0 to 1 in $l \times n$ dimensions. The calculation process is as follows:
\begin{gather}
M(\boldsymbol{r}) = mean (Mask([\hat{\boldsymbol{r}}_{i}]_{1 \leq i \leq T})) \\
\boldsymbol{Q} = \sigma (MLP(M(\boldsymbol{r}) ))
\end{gather}
where $\boldsymbol{Q}=[q_{ij}]_{1 \leq i \leq m}^{1 \leq j \leq n} \in \mathbb{R}^{m \times n}$ denotes the decoded similarity distribution. $mean(\cdot)$ denotes mean pooling operation. $\sigma$ denotes the sigmoid function. By using two sets of decoder parameters, we can decode the two modal interest sequences $\hat{\boldsymbol{r}}^{m_1}$ and $\hat{\boldsymbol{r}}^{m_2}$ into $\boldsymbol{Q}^{m_1}$ and $\boldsymbol{Q}^{m_2}$, respectively. The Kullback-Leibler divergence is used to constrain the decoded distribution to be close to the corresponding true distribution, thereby reducing the information loss:
\begin{gather}
\begin{aligned}
\mathcal{L}_{dec} &= D_{KL}(\boldsymbol{P}^{m_1}||\boldsymbol{Q}^{m_1}) + D_{KL}(\boldsymbol{P}^{m_2}||\boldsymbol{Q}^{m_2})\\
&=\mathop{\mathbb{E}} [ \begin{matrix} \sum_{i=1}^m \sum_{j=1}^n p_{ij}^{m_1}log \frac{p_{ij}^{m_1}}{q_{ij}^{m_1}}  \end{matrix} + \begin{matrix} \sum_{i=1}^m \sum_{j=1}^n p_{ij}^{m_2}log \frac{p_{ij}^{m_2}}{q_{ij}^{m_1}}  \end{matrix}]
\end{aligned}
\end{gather}

\subsection{Model Learning} 
Our main objective is to predict user click probability on the target item accurately. To achieve this, we must pool the multimodal interest sequence, integrate it with other feature vectors, and process it using a deep network.  Since the focus of this paper is on the adaptive fusion of multimodal interest, we use a commonly used compressed sequence method, such as the Deep Interest Network (DIN) \cite{din}, as the backbone. Ultimately, all embedding vectors and the user interest representation vector are merged and fed into an MLP module to generate the final prediction as follows:
\begin{gather}
\boldsymbol{h} = DIN( [\boldsymbol{v}_1, ..., \boldsymbol{v}_j, ..., \boldsymbol{v}_T]) \\
\hat{y}= \sigma(DNN(\boldsymbol{v}_u \oplus  \boldsymbol{v}_i \oplus \boldsymbol{h} \oplus  \boldsymbol{r}_{*}^{m_1} \oplus  \boldsymbol{r}_{*}^{m_2}))
\end{gather}
Then, the cross entropy loss is used to optimize model parameters:
\begin{gather}
\mathcal{L} = -\frac{1}{N} \begin{matrix} \sum_{i=1}^N y_i \log\hat{y}_i + (1-y_i) \log(1-\hat{y}_i) \end{matrix}
\end{gather} 
where $N$ is the total number of training instances. The overall objective function of DAME is derived by
\begin{gather}
\mathcal{L}_{DAME} := \mathcal{L} +   \lambda_{dec} \mathcal{L}_{dec}
\end{gather}
where $\lambda_{dec}$ regulates the significance of the decoding term.

\section{Experiments}
This section comprehensively describes our experiments, including the dataset information, model experiment setup, comparative analysis of the DMAE's performance with existing CTR prediction methods, and compatibility and ablation experiments. Our aim is to answer the following questions in this section:  \\
\textbf{$\bullet$ RQ1:} Does the DMAE model outperform existing recommendation methods in terms of predictive performance?  \\
\textbf{$\bullet$ RQ2:} What is the impact of each module in the DMAE model? \\
\textbf{$\bullet$ RQ3:} What is the influence of each hyper-parameter in DMAE? \\
\textbf{$\bullet$ RQ4:} Can the DAME model alleviate the cold-start problem?  \\ 

\begin{table}[t]
	\small
	\centering
	\begin{tabular}{ccccc}
    		\hline Dataset & \# User & \# Item & \# Categories & \# Impressions \\
		\hline Amazon-Books & 75,053 & 358,367 & 1,583 & 150,016  \\
		Amazon-Electro & 192,403 & 63,001 & 801 & 1,689,188  \\
		MovieLens. & 138,493 & 27,278 &21& 20,000,263 \\
		Industrial. & 10,125,327 & 1,652,178 &3,949 & 873,945,605 \\
		\hline
	\end{tabular}
	\caption{Statistics of four benchmark datasets.}
	\label{datasets}
\end{table} 

\subsection{Datasets}
This paper evaluates the predictive performance of different methods on four datasets. Table \ref{datasets} presents detailed statistics. \\
\textbf{$\bullet$ Amazon-Book  \footnote{ http://jmcauley.ucsd.edu/data/amazon/}}.  It consists of item reviews and metadata from Amazon. We use the Books subset of the Amazon dataset, which contains 75053 users, 358367 items, and 1583 categories \cite{sim}. For this dataset, we consider reviews as one kind of interaction behaviors. The modality information includes text and images. \\
\textbf{$\bullet$ Amazon-Electro}.  It's also from the Amazon dataset and also has text and images data. It contains 192,403 users, 63,001 goods, 801 categories and 1,689,188 samples. User behaviors in this dataset are rich, with more than 5 reviews for each users and items. \\
\textbf{$\bullet$ MovieLens  \footnote{ https://grouplens.org/datasets/movielens/20m/}}. It contains 138,493 users, 27,278 movies, 21 categories and 20,000,263 samples. Following \cite{din}, we transform it into a binary classification data to make it suitable for prediction task, and we predict whether user will rate a given movie to be above 3(positive label) based on historical behaviors. To acquire texts and images data, we crawl multimodal information from IMDB \footnote{ https://www.imdb.com/}. \\
\textbf{$\bullet$ Industrial Dataset}. The dataset is obtained from an international e-commerce platform's online display advertising system. The training set is composed of samples from the last 20 days, while the test set consists of exposure samples from the subsequent day, which is a common practice in industrial modeling. The modality information also includes text and images.

\begin{table*}[t]
	\caption{Prediction performance on datasets of CTR prediction models. $\Delta_{AUC} $ and $\Delta_{Logloss} $ are calculated to indicate averaged performance boost compared with the baseline (DIN) over datasets. * indicates p-value < 0.05 in the significance test.}
	\begin{tabular}{c|cc|cc|cc|ccc|cc}
    \toprule
       \multirow{2}{*}{\diagbox{Method}{Dataset}}& \multicolumn{2}{c|}{Amazon-Books} & \multicolumn{2}{c|}{Amazon-Electro} &\multicolumn{2}{c|}{MovieLens}& \multicolumn{3}{c|}{Industrial.}&$\Delta_{AUC}$&$\Delta_{Logloss} $ \cr
   \cmidrule(lr){2-10}
   & AUC& Logloss& AUC& Logloss& AUC& Logloss& AUC&$GAUC_{pv}$&{Logloss} & $\uparrow$ &$\downarrow$ \cr 
    \midrule
    Wide\&Deep &0.6630 &0.6403 &0.8637 &0.3672   &0.7212 &0.6092 &0.6990  & 0.6070 & 0.1062 &-3.95\%  & +0.0170 \cr
	FRNet &0.6813  &0.6275  &0.8731 &0.3562  &0.7241 &0.6053  &0.7027 &0.6093  &0.1052  &-2.82\%  & +0.0099 \cr
	Average &0.7248 &0.6106  &0.8797 &0.3504  &0.7329 &0.6012 &0.7066 & 0.6106 & 0.1049 & -0.71\%  & +0.0031  \cr
	GRU4Rec  &0.7238 &0.6112  &0.8782 & 0.3526 &0.7331  &0.6005 &0.7108 & 0.6089 &0.1044 &-0.63\%  & +0.0035 \cr
	DIN&0.7281  &0.6096  &0.8857 &0.3416  &0.7366 &0.5993    & 0.7152 & 0.6097 &0.1042&-  & - \cr
	BST&0.7264 &0.6109  &0.8824  &0.3401 &0.7343 &0.6001   & 0.7155 &0.6105 &0.1041 &-0.22\%  & +0.0002 \cr
	\midrule
	UniSRec&0.7390  &0.6026  &0.8946 &0.3392  &0.7409 &0.5909    & 0.7188 & 0.6121 &0.1039& +0.89\%  & -0.0045 \cr
	LRD&0.7371  &0.6031  &0.8928 &0.3402  &0.7395 &0.5921    & 0.7204 & 0.6136 &0.1038& +0.79\%  &  -0.0039 \cr
	DIMO&0.7501  &0.5961  &0.8972 &0.3378  &0.7431 &0.5887    & 0.7233 & 0.6143 &0.1036& +1.58\%  &  -0.0071 \cr
	SimTier&0.7526  &0.5946  &0.9034 &0.3355  &0.7482 &0.5846    & 0.7252 & 0.6181&0.1033& +2.08\%  &  -0.0092\cr
	\midrule
    DMAE&\textbf{0.7645*} & \textbf{0.5873*} & \textbf{0.9105*} & \textbf{0.3334*} & \textbf{0.7570*} &  \textbf{0.5793*} &\textbf{0.7319*} &\textbf{0.6249*} &\textbf{0.1028*} & \textbf{+3.23\%} & \textbf{-0.0129} \cr
    \bottomrule
    \end{tabular}
	\label{results}
	\vspace{-0.1cm}
\end{table*} 

\begin{table*}[t]
	\caption{The results of ablation study. We record the mean results over 5 runs. Std $ \leq$ 0.02\%. $\Delta_{AUC} $ and $\Delta_{Logloss} $ are calculated to indicate averaged performance degradation compared with the original method (DMAE) over the four datasets.}
	\begin{tabular}{c|cc|cc|cc|ccc|cc}
    \toprule
     \multirow{2}{*}{\diagbox{Method}{Dataset}}& \multicolumn{2}{c|}{Amazon-Books} & \multicolumn{2}{c|}{Amazon-Electro} &\multicolumn{2}{c|}{MovieLens}& \multicolumn{3}{c|}{Industrial.}&$\Delta_{AUC}$&$\Delta_{Logloss} $ \cr
   \cmidrule(lr){2-10}
   & AUC& Logloss& AUC& Logloss& AUC& Logloss& AUC&$GAUC_{pv}$&{Logloss} & $\uparrow$ &$\downarrow$ \cr 
    \midrule
    w/o MIEU-se &0.7582 &0.5898 &0.9069 &0.3340 &0.7541 &0.5814 & 0.7287 &0.6218  &0.1030 &-0.51\% &+0.0013 \\
		w/o MIEU-t &0.7580 &0.5901 &0.9072 & 0.3339 &0.7555 &0.5806 &0.7279 &0.6213 &0.1031 &-0.48\% &+0.0012 \\
		 w/o MIFU &0.7539 &0.5924 &0.9042 &0.3347 &0.7501 &0.5839 &0.7261 &0.6192 &0.1032 &-0.94\% &+0.0028  \\
		 w/o IDDU &0.7561 &0.5909 &0.9051 &0.3342 &0.7527 &0.5827 &0.7273 &0.6201 &0.1031 &-0.72\% &+0.0020  \\
	\midrule
    DMAE &\textbf{0.7645*} & \textbf{0.5873*} & \textbf{0.9105*} & \textbf{0.3334*} & \textbf{0.7570*} &  \textbf{0.5793*} &\textbf{0.7319*} &\textbf{0.6249*} &\textbf{0.1028*} & - & - \\
    \bottomrule
    \end{tabular}
\label{ablation}
\vspace{-0.3cm}
\end{table*} 

\subsection{Competitors and experiment settings}
Models under comparison and their relevant details are as follows:
\subsubsection{ID-based methods} Recommendation model that relies solely on sparse ID features to learn co-occurrence of items. \\
\textbf{$\bullet$ Wide\&Deep} \cite{widedeep}.  It jointly trains a linear and a deep neural model, which combines the benefits of memorization and generalization. \\
\textbf{$\bullet$ FRNet} \cite{frnet}.  It can be used to learn context-aware feature representations in CTR prediction models. We apply FRNet into FM, which has fewer parameters than high-order or ensemble methods. \\
\textbf{$\bullet$ Average Pooling} \cite{rec1}.  It averages all item embeddings in the sequence and inputs them into the MLP together with other features. \\
\textbf{$\bullet$ GRU4Rec}\cite{gru4rec}.  It is a session-based framework that uses the RNN structure to capture sequential patterns in each session of users. \\
\textbf{$\bullet$ DIN}\cite{din}.  It utilizes the attention mechanism to activate relevant users' behaviors and learns the representation for users' interests. \\
\textbf{$\bullet$ BST}\cite{bst}.  It uses the powerful Transformer model to capture the sequential signals underlying users’ behavior sequences. 

\subsubsection{Multimodality-based methods} Recommendation model that enhances the ability to model user interests by introducing multimodal information of item. \\
\textbf{$\bullet$ UniSRec}\cite{unisrec}. It utilizes the associated description text of items to learn transferable representations across different scenarios. \\
\textbf{$\bullet$ LRD}\cite{lrd}. It proposes to discover latent relations based on LLM for relation-aware sequential recommender systems. We set DIN as the backbone of LRD in this paper. \\
\textbf{$\bullet$ DIMO}\cite{dimo}. It presents a multi-view self-supervised disentanglement to disentangle ID and modality effects. \\
\textbf{$\bullet$ SimTier}\cite{scl}. It converts a set of high-dimensional multimodal representations into the representations of the user’s interest. 

\textbf{Experiment Settings}. We employ the identical experimental setup as previous studies to ensure a fair comparison of experimental results \cite{din, sim}. We harness the GPT-3 and EVA-02 \cite{eva} to obtain text and image knowledge representations of items, respectively.  All models are implemented in the Tensorflow framework and trained using the Adam optimizer \cite{adam}. We apply exponential decay with an initial learning rate of 0.001 and a decay rate of 0.9. In the public datasets, layers of fully connected network are set by $200 \times 80 \times 1$, and we perform grid search for the embedding dimension in $\left\{4, 8, 16, 32 \right\}$. In the industrial dataset, layers of fully connected network are set by $512 \times 256 \times 64 \times 1$, with embedding dimensions of ID features set at 8. We set the window size in MIFU to be consistent with the time interval $n$ in Section 4.3.1.

\subsection{Performance Comparison (RQ1)}
We evaluate the model's performance using two metrics, AUC and Logloss. Moreover, recognizing the significance of ranking performance under a single user request (pv) in real advertising systems, we calculate the group AUC \cite{gauc} at the request level, named as $GAUC_{pv}$, on the industrial dataset. Optimal performance is highlighted in bold. The experimental results reveal that DMAE can achieve the best prediction performance comparable to all baseline methods on all datasets. This finding provides evidence of the effectiveness of DMAE for user multimodal interest modeling.

From Table \ref{results}, it is evident that all methods incorporating multimodal information outperform ID-based methods. This result is easily comprehensible. Recommendation systems frequently encounter the issue of sparse data. Incorporating explicit multimodal information can yield substantial information gain, assisting the model in better capturing the relationship between items and, in turn, more accurately modeling user interests. Nevertheless, it is evident that the performance of the UniSRec and LRD models is noticeably inferior to that of the DIMO and SimTier methods, which disentangle the multimodal representations and ID embedding vectors.  This is due to the fact that imposing constraints on the embedding vectors associated with the IDs to capture multimodal information not only results in the loss of multimodal information, but also disrupts the original co-occurrence significance of the ID embedding vectors, consequently diminishing the information gain introduced by the multimodal information. Our method, DMAE, surpasses the aforementioned multimodal modeling methods, further affirming the importance of multimodal adaptive fusion.

Moreover, as observed in Table \ref{results}, Wide\&Deep and FRNet exhibit the lowest performance. While FRNet and CAN improve the model's capacity for feature interaction through structural refinements, the absence of user sequence data inherently limits information throughput, posing a barrier to achieving peak performance. This underscores the significance of incorporating user sequence modeling in recommendation models. At the same time, although BST introduced Transformer to enhance the model's ability of modeling user sequences, it also resulted in an exponential increase in computation, making it impractical for online service systems. Furthermore, the performance of BST on four datasets is not superior to that of DIN. This is due to that BST requires lots of samples to converge the large number of parameters. This is why we use DIN as the backbone of the multimodal framework in this paper.

\subsection{Ablation Study (RQ2)}
To investigate the contribution of each module in DAME, we conduct ablation studies on datasets. We have four variants as follows: \\
\textbf{$\bullet$ w/o MIEU-se} removes the design of similarity embedding, which means that DMAE will adopt the bucketing strategy. \\
\textbf{$\bullet$ w/o MIEU-t} removes the timing dimension from the user interest distribution, which means that the interest distribution is constructed solely based on the similarity scores in the sequence. \\
\textbf{$\bullet$ w/o MIFU} removes the multimodal interest fusion unit from DMAE, which means that we directly pool two multimodal interest representation sequences and use them for final predictions. \\
\textbf{$\bullet$ w/o IDDU} removes the decoding unit from DMAE, which means that only the final prediction loss is used for model learning.

Table \ref{ablation} displays the results of the ablation experiment, demonstrating that all variants perform inferiorly to DMAE, providing strong evidence that each module of DMAE contributes to the final prediction. In particular, the most significant performance degradation is observed when the Multimodal Interest Fusion Unit is removed (w/o MIFU), signifying the crucial importance of this unit in multimodal interest modeling. This also validates our perspective that multimodal cross fusion should be based on the user behavior sequence, allowing us to capture inter-modal and intra-modal interaction information, leading to a more comprehensive utilization of multimodal information. Simultaneously, the removal of the Interest-Distribution Decoding Unit (w/o IDDU) results in performance degradation. We posit that removing the decoding unit implies that the model can solely depend on the final prediction loss for learning, potentially jeopardizing the ability of encoded user interest representation under different modalities to accurately represent corresponding modality information. Therefore, we assert that IDDU plays a crucial role in multimodal modeling, as it influences the model's learning of multimodal interest representations, which forms the basis of our model. Together with the encoder, it determines the capacity of the multimodal model. Similarly, we observe that modifying the encoding of the similarity score in MIEU (w/o MIEU-se) results in a substantial decrease in model performance. This is due to the encoding of the similarity score setting the threshold for incorporating multimodal information. Inaccurate representation of the similarity score by our designed encoding framework will impede the model's capability. For example, if we use the same representation vectors to depict different levels of user interest (a basic bucketing strategy will lead to this issue), our subsequent multimodal cross unit and decoding structure will not function properly, as the multimodal information will essentially be lost during the encoding process. This further emphasizes the importance of our investigation into an appropriate encoding structure. At last, excluding the time feature during encoding (w/o MIEU-t) also results in a substantial decrease in model performance, as shown in Table \ref{ablation}. This is due to the fact that user interests evolve over time, and the removal of the time feature may impede our capability to model the dynamic changes in user interests. 

\subsection{Hyper-Parameters Sensitivity Analysis (Q3)}
\begin{figure}[t]
  \centering
  \includegraphics[width=\linewidth]{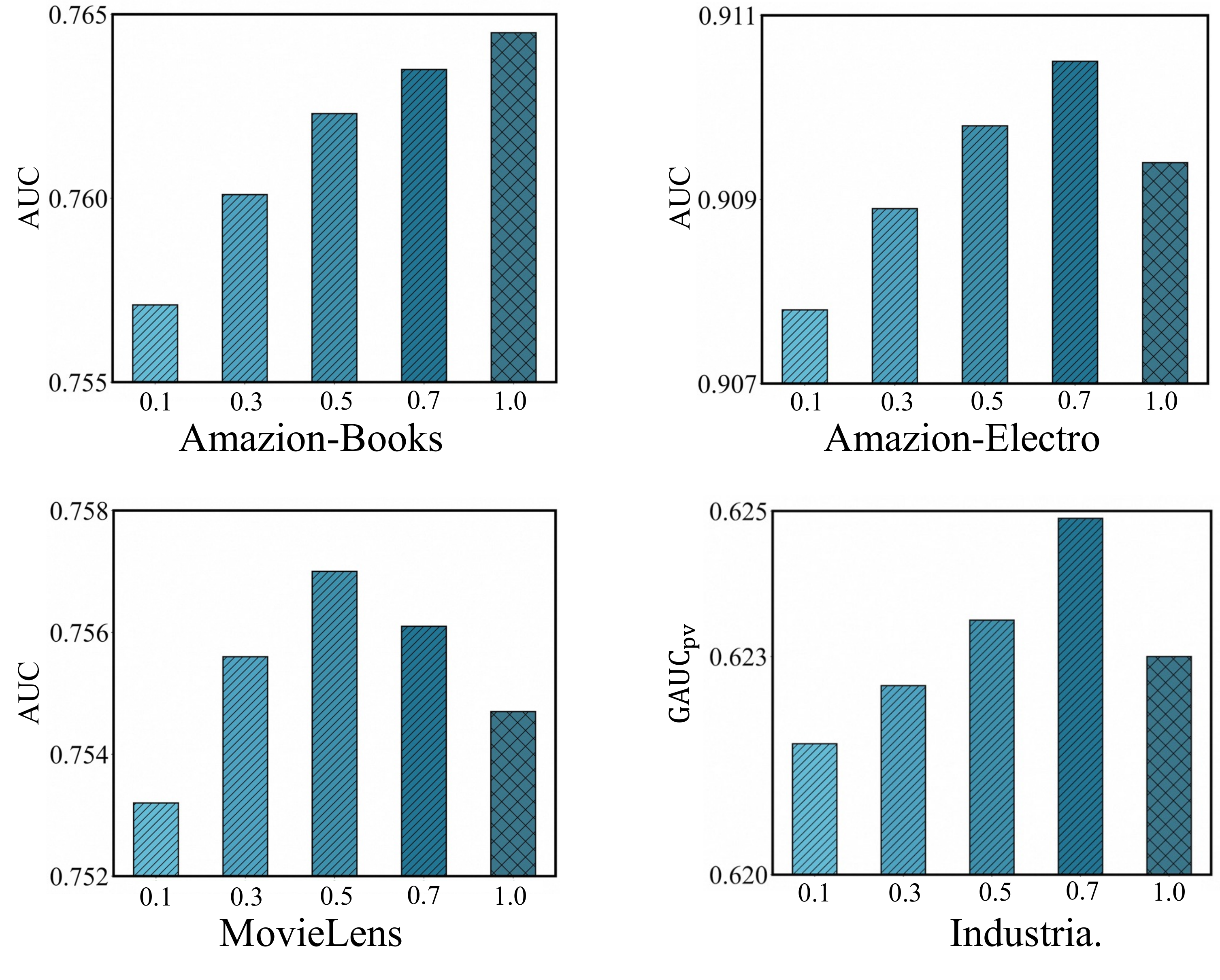}
  \caption{The performance of DMAE under different predefined coefficient of the decoding loss on four datasets.}
  \label{hyper}
\end{figure}

This section aims to investigate the impact of hyper-parameters of our model. The above ablation experiment shows that the IDDU in DMAE significantly affects the model performance. Therefore, two hyper-parameters of IDDU are considered: the decoding loss coefficient $\lambda_{dec}$ and the size of $l$ and $n$ for creating the user interest distribution label. Since the goal of our model is to predict whether the user will click on the target item, we aim to focus DMAE's attention on the prediction loss. Therefore, we conduct experiments with $\lambda_{dec}$ in the range of $\left\{0.1, 0.3, 0.5, 0.7, 1.0 \right\}$.  The experiment results are depicted in Figure \ref{hyper}. It is observed that DMAE achieves the best performance when $\lambda_{dec}$ is set to 1.0, 0.7, 0.5, and 0.7 for four datasets, respectively. This is believed to be due to the fact that a smaller coefficient may not sufficiently impact the model learning, and conversely, a larger coefficient poses the risk of diverting the optimization direction from the primary prediction task.  

What's more, we conduct parameter adjustment experiments of $l$ and $n$ on the industrial dataset. As the length of user sequences in the dataset is truncated to 64, we conducted experiments with values of $l$ and $n$ in the range of $\left\{1, 5, 10, 32, 64\right\}$ and $\left\{1, 5, 10, 50, 100\right\}$, respesctively. From the Table \ref{hyper2}, we can find that DMAE achieves the best performance when the both values are set to 10. This result is readily comprehensible. A large time interval, such as setting $l$ to $1$, neglects the time factor in interest distribution construction, resulting in poor constraint effect of the decoding loss. Conversely, a small time interval or a similarity interval imposes high learning requirements on the model parameters for the decoding loss, causing difficulty in model convergence. Hence, a moderate interval is necessary for constructing the user interest distribution.

\subsection{Online A/B Testing Results (Q4)}
Due to the model scoring bias affecting the penetration opportunities of new items (i.e., cold-start items), it is challenging to assess the model's impact on new items through offline experiments. Therefore, we conduct online A/B advertising traffic tests on DMAE and focus on the following indicators: 1). The impression proportion of new items in the overall traffic exposure $IMP_{new}$, as we aim for greater penetration opportunities for new items; 2). Overall Click-Through Rate (CTR) of new items exposure $CTR_{new}$, to identify potential popular new items; 3. Overall CTR of the advertising traffic and revenue (REV) changes, aiming to improve overall advertising efficiency by exposing more high-quality items.

\begin{table}[t]
	\caption{The GAUC results of adjusting $l$ and $n$. }
	\begin{tabular}{c|c|c|c|c|c}
    \toprule
    {\diagbox[height=4ex,width=4em]{$n$}{$l$}} & 1&5& 10& 32& 64 \cr 
    \midrule
    		1 &0.7562 &0.7574 &0.7583 &0.7531 &0.7504\\
      \midrule
		5  &0.7571 &0.7608 &0.7624 &0.7586 &0.7531\\
		  \midrule
		10  &0.7580 &0.7618 &0.7645&0.7602 &0.7553\\
		  \midrule
		50  &0.7572&0.7598 &0.7613 &0.7561 &0.7512\\
		  \midrule
		 100  &0.7551 &0.7563 &0.7574 &0.7505 &0.7462 \\
    \bottomrule
    \end{tabular}
\label{hyper2}
\end{table}

Specifically, we conducted a 7-day A/B test on our e-commerce platform using the DMAE. The base model is DIN, which does not utilize multimodal information. We define new items as those that have not appeared in the past 20 days, aligning with the training sample date of our model. According to the experimental results, DMAE increased $IMP_{new}$ by 9.25\% compared to the base model, significantly improving the exposure opportunities for new items and aiding in their rapid growth. Additionally, $CTR_{new}$ increased by 14.83\%, suggesting the ability of DMAE to recommend more suitable new items to users. Ultimately, the overall CTR increased by 12.9\% and revenue (REV) increased by 11.63\%, which shows that we also improved the performance of overall recommendations.

\textbf{Online Deployment}. Since the modules for multimodal interests of DMAE and DIN process independent inputs, their computations can be parallelized on GPUs. DMAE introduces sliding window attention and retains the time complexity as O(L), similar to DIN. Consequently, while DMAE requires additional storage for multimodal vectors, it maintains comparable online response times to DIN. To save the storage, quantization \cite{int8} was utilized during online deployment to reduce memory space by 40\% as a result of the extensive storage space needed for multimodal representation vectors, with only a minor impact on online performance.

\section{Conclusions}
This paper proposes the Distribution-Guided Multimodal-Interest Auto-Encoder (DMAE) to dynamically fuses users' multimodal interest distributions on the target item based on historical click sequences. Specifically, DMAE employs an auto-encoder architecture that: (1) encodes users' interest intensity from the representation space of the item similarity distribution under different modalities, and (2) incorporates a decoding unit that constrains the encoded user interest representations to restore the real interest distribution, thereby restricting the model parameter learning and minimizing the loss of multimodal information during the encoding process. Meanwhile, to better capture user multimodal interests, DMAE performs cross-modal and intra-modal fusion on encoded interest representations, enabling behavior-level multimodal interest adjustment and dynamic fusion. Finally, our experiments provide substantial evidence for the superiority of DMAE.

\end{document}